\documentclass[conference]{IEEEtran}
\usepackage{booktabs} 
\usepackage{graphicx}
\usepackage{subcaption}
\usepackage{multirow}
\usepackage{algorithm2e}
\usepackage{tabularx}
\usepackage{tabulary}
\usepackage{cleveref}
\usepackage{float}
\usepackage{amsmath}
\usepackage{makecell}
\usepackage{color}
\renewcommand\baselinestretch{0.93}

\ifCLASSINFOpdf
\else
\fi

\begin{document}
%
\title{Scalable NoC-based Neuromorphic Hardware Learning and Inference}

\author{\IEEEauthorblockN{Haowen Fang$^{1}$, Amar Shrestha$^{1}$, De Ma$^{2}$, Qinru Qiu$^{1}$}
\IEEEauthorblockA{$^{1}$Department of Engineering and Computer Science, Syracuse University, Syracuse, New York\\
$^{2}$College of Computer Science and Technology, Zhejiang University, China\\
Email: { $^1$\{hfang02,amshrest,qiqiu\}@syr.edu,$^2$\@made@hdu.edu.cn} }
}


%


\maketitle

\begin{abstract}
Bio-inspired neuromorphic hardware is a research direction to approach brain's computational power and energy efficiency. Spiking neural networks (SNN) encode information as sparsely distributed spike trains and employ spike-timing-dependent plasticity (STDP) mechanism for learning. Existing hardware implementations of SNN are limited in scale or do not have in-hardware learning capability. In this work, we propose a low-cost scalable Network-on-Chip (NoC) based SNN hardware architecture with fully distributed in-hardware STDP learning capability. All hardware neurons work in parallel and communicate through the NoC. This enables chip-level interconnection, scalability and reconfigurability necessary for deploying different applications. The hardware is applied to learn MNIST digits as an evaluation of its learning capability. We explore the design space to study the trade-offs between speed, area and energy. How to use this procedure to find optimal architecture configuration is also discussed.

\end{abstract}

\begin{IEEEkeywords}
Spiking neural network, Network on chip, STDP learning, Unsupervised learning
\end{IEEEkeywords}

%
\IEEEpeerreviewmaketitle

\section{Introduction}\label{sec:intro}

In the field of deep learning, convolutional neural networks (CNNs) and recurrent neural networks (RNNs)  are
developed to perform a series of human-level cognitive applications \cite{li2016assisting} \cite{ren2017sc}. However, the tremendous computation and memory requirement have been seriously challenging the processing efficiency of deep learning systems \cite{wang2018towards} \cite{lin2017fft}. The limitations of Von Neumann architecture coupled with increasing power demands due to Dennard scaling and the approaching end of Moore's Law have motivated multiple research efforts into low-power, highly parallel and distributed computing architecture \cite{wang2018c} \cite{furber2013overview} \cite{ding2017c} \cite{li2017hardware} and brain-inspired computing architecture \cite{merolla2014million} \cite{liprobabilistic}. Brain as a source of inspiration is not surprising given its ability to process massive amounts of real-time information while consuming less than 20 W of power \cite{ananthanarayanan2009cat}. The goal of neuromorphic hardware design is to explore the bio-inspired architecture to achieve cognitive functions in real time utilizing lower power and smaller footprint than the traditional Von Neumann architectures.

Brain, in simplistic terms, is a collection of neurons, interconnected in a vast network through links called synapses. The communication between neurons in the vast network provides the brain its processing abilities and perform pattern recognition, classification, associative memory, reasoning etc. The basis of this communication is the short asynchronous electrical pulses/action potentials called spikes. Spiking neural networks (SNNs), which use spikes as the basis for communication, is the third generation of neural networks \cite{maass1997networks}. As each neuron works asynchronously in an event-driven manner, SNNs have the potential to reach very low energy dissipation.  When spiking activity in SNNs is stochastic i.e. spikes are generated as a stochastic process,  the information is carried by the statistics of a group of spikes instead of individual spikes. This makes the SNN more biologically plausible and also improves its fault tolerance and noise (delay) resilience.

The network of neurons in the brain can learn patterns by modifying the synapses linking the neurons based on their causal relative spike timings. This local and causal learning rule is called Spike-Timing Dependent Plasticity ( STDP )\cite{sjostrom2010spike}. As it is based only on local information of individual neurons, fully distributed learning\cite{sjostrom2010spike} can be achieved on SNNs. Several challenges exist in implementing STDP learning on hardware. Firstly, the STDP rule is typically an exponential function, which is expensive for hardware implementation. Secondly, since the final value of synaptic weight is unknown during the learning, the hardware implementation should consider the worst case and be ready to provide a wide range and high precision for every synapse. Hence, more memory is required for hardware neurons that has learning capability than the hardware neurons that performs inference only.

SNN hardware requires massive interconnection for parallelism and scalability. The Network-on-Chip(NoC) architecture has been used to provide on-chip communication for massive parallel systems. Traditional NoC design aims at minimizing communication latency and router congestion. As we will show later, due to time-multiplexed nature in neuron hardware implementation, and asynchronous and stochastic neuron behavior, latency of inter-neuron communication is not a performance bottleneck in hardware SNN. This property enables us to significantly simply router design and reduces hardware cost.

In this work, we adopt Q2PS approximation of STDP rule \cite{shrestha2017stable}  to simplify the hardware exponential function. Different synaptic weights are encoded differently to provide both a wide range and a high precision without increasing the storage. A very low-cost NoC is designed to provide just enough communication capability for an SNN. Our main contributions include:

\begin{enumerate}
	\item A low-cost time-multiplexed hardware spiking neuron design. Replacing multiplication and exponential functions with add and shift operations reduces hardware complexity as well as power consumption. The time-multiplexed physical neuron design improves resource utilization and neuron density.
	\item A compact NoC design with a low-cost router, which is application specific and optimized for spiking neural network.
	\item Experimental demonstration of STDP learning capability of the hardware by applying it to unsupervised learning of MNIST digits.
	\item Design space exploration to study speed, area and energy trade-offs and suggestions of design choices based on the analysis.
\end{enumerate}

\section{Related Works} \label{related_works}
There have been several existing research works on SNN from the hardware perspective  \cite{furber2013overview} \cite{merolla2014million} \cite{liprobabilistic}. The IBMs TrueNorth neurosynaptic processor \cite{merolla2014million} has achieved state-of the-art performance with minimal energy footprint on many tasks \cite{shrestha2017spike}\cite{andreou2016real}, but it does not provide in-hardware learning capabilities. SpiNNaker \cite{furber2013overview} has also been popular in the research community as a testbed for SNN applications. SpiNNaker Chip Multiprocessor integrates 18 ARM cores. It is capable of massively parallel simulations for spiking neural network. \cite{yao2017face} presents an analogue device to implement artificial synapse with high energy efficiency, which shows 30 nJ energy consumption for an epoch of classification task. \cite{azghadi2013programmable} proposed a programmable CMOS neuromorphic chip. The architecture aims at implementing biologically plausible circuits and is limited in scalability. To address the scalability issue, there are works adopt NoC technique with SNN. EMBRACE is an FPGA based flexible and reconfigurable SNN architecture\cite{cawley2011hardware}. It uses NoC to handle inter-neuron communication. EMBRACE also features a genetic based onchip training. It randomly initializes neuron configurations and performs fitness evaluation, crossover and selection until the optimal SNN configuration is obtained. \cite{carrillo2013scalable} presents H-NoC, an architecture for spiking neural network. The goal of H-NoC is to reduce packet delay and it assumes that each neuron in the SNN has a dedicated port to the router although the detail of the neuron is not given. As we will show later, the time multiplexed neuron core design and asynchronous nature of the neuron activities relax the latency constraint. A more simplified NoC design suffices the SNN application.

Hardware implementations of STDP learning \cite{park2013neuromorphic} \cite{cruz2012energy} focus more on circuit and device level analysis to achieve variable synaptic plasticity instead of scalability. \cite{ahmed2016system} proposed a digital hardware neuron model for synaptic plasticity, it focuses on the design of individual neuron cores, interconnection and scalability are not addressed. Few analog VLSI approaches of synaptic plasticity are proposed in \cite{schemmel2006implementing}\cite{huayaney2016vlsi}\cite{indiveri2006vlsi}, which focus on the individual synapses design without addressing large scale network implementation and architectural design. Emerging memristive devices have also been studied to realize artificial synapse and synaptic plasticity\cite{yuan2017memristor}\cite{deng2015complex}\cite{covi2016analog}\cite{prezioso2016self}\cite{saighi2015plasticity}. However these researches are still at proof-of-concept level and the fabrication technology of memristive device is not yet mature.

\section{Neuron Model and Learning Rule} \label{stdp_rules}

\begin{figure}
  \centering
  \includegraphics[width=0.4\linewidth]{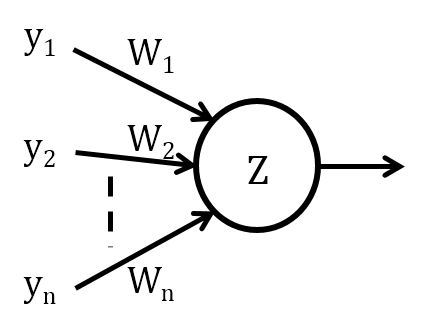}
  \caption{Generic neuron model.}
  \label{fig:neuron_model}
\end{figure}

We utilize a simplified version \cite{ahmed2016probabilistic} of the neuron model proposed in \cite{nessler2013bayesian}. Here the membrane potential $u(t)$ of neuron $Z$ is computed as
\begin{equation} \label{eq:bayesian}
  u(t) = w_0 + u(t-1) + \sum_{i=1}^{n} w_i \cdot y_i(t)
\end{equation}
where $w_i$ is the weight of the synapse connecting Z to its $i_{th}$ pre-synaptic neuron $y_i$, $y_i(t)$ is 1 if $y_i$ issues a spike at time t, and $w_0$ models the intrinsic excitability (bias) of the neuron Z. An integrate and fire neuron Z spikes when the membrane potential crosses the threshold and then its membrane potential is reset to 0. When the threshold is set to be random over a specified range, the stochastic integrate-and-fire neuron (SIF) approximates the Bayesian neuron in \cite{ahmed2016probabilistic}. 

In order to aggregate or relay spike activities, we also introduce spiking Rectified Linear Unit (ReLU) neuron. A ReLU neuron accumulates every weighted input spike and discharges it over time resembling a burst firing pattern. After a spike, the membrane potential of a ReLU neuron is computed as:

\begin{equation} \label{eq:relu}
u(t) = u(t - 1) + \sum_{i=1}^{n} y_i(t) - U_{th}
\end{equation}

STDP is the basis for learning in a spiking neuron model. Multiplicative STDP are stable but induces low competition whereas additive rules are highly competitive but unstable. Both qualities, stability and competitiveness, are highly desirable. Most existing STDP rules utilize the exponential function, which is expensive for digital hardware implementation. Here, we utilize the low-cost Q2PS STDP rule proposed in \cite{shrestha2017stable} to approximate the exponential and multiplications using shifters, adders and a priority encoder. The analysis in \cite{shrestha2017stable} shows that the Q2PS rule is stable and highly competitive. The rule is given in Equation \label{eq:quantization}:

\begin{equation} \label{eq:quantization}
  \Delta w_i = \begin{cases}
1 << {|\bar{Q}|},  & \text{$if \bar{Q} > 0$}\\
1 >> {|\bar{Q}|},  & \text{$if \bar{Q} < 0$}
\end{cases}
\end{equation}

Where  $\bar{Q}$ is the quantization of $Q$ through priority encoding which is given as below.

If $t_{post} - t_{pre} < \tau_{LTP}$, then
\begin{equation} \label{eq:ltp}
  Q = \eta'_{LTP} - w_i
\end{equation}

If $t_{post} - t_{pre} > \tau_{LTP}$ or $t_{pre} - t_{post} < \tau_{LTD}$, then
\begin{equation} \label{eq:ltd}
  Q = \eta'_{LTD} + w_i
\end{equation}

Where  $\eta'_{LTP} = \log_2{\eta_{LTP}}$ and $\eta'_{LTD} = \log_2{\eta_{LTD}}$. And $t_{post}$ and $t_{pre}$ are the time steps at which the pre and post-synaptic neuron spikes,  $\tau_{LTP}$ and $\tau_{LTD}$  are the LTP and LTD window and $\eta_{LTP}$ and $\eta_{LTD}$ are the LTP and LTD learning rates respectively.

The base 2 exponential function in Q2PS can be implemented using a barrel shifter with very low hardware cost. The weight learned by this rule has a limited range\cite{shrestha2017stable}, which will be explored to reduce the storage requirement, as will be discussed in section \ref{sec:hardware_design}.

\section{Hardware architecture}\label{sec:hardware}

The proposed architecture consists of a grid of homogeneous neurons. Each individual neuron's behavior is programmable and detailed neuron configuration will be discussed in Section \ref{sec:operation}. We adopt a globally asynchronous, locally synchronous (GALS) approach and avoid using a global clock. Neurons and routers work asynchronously in different clock domains. NoC is used as the global communication infrastructure to address massive interconnections of SNN. In this section, we will discuss the NoC design, the router architecture, the network interface and the hardware neuron design.

\subsection{Network-on-Chip design}\label{sec:noc}
SNNs have massive numbers of interconnected neurons running simultaneously with each neuron having fan-outs larger than $10^3$ \cite{pakkenberg2003aging}. Traditional on-chip communication solutions such as bus or point-to-point connection are limited in either scalability or flexibility \cite{jantsch2003networks}. NoC has been widely used to provide inter-core communication for massive parallel on-chip systems. A typical NoC architecture consists of three components; router, channel and PE (processing element). Routers are interconnected by channels. Each PE is attached with a router and communication with each other via multi-hop packet transmission. Based on the destination address provided by the packet, routers make routing decision to forward it either to the next router or to the local PE. In this way, arbitrary network topology can be implemented.

Traditional NoC design aims at minimizing the communication latency and router congestion to ensure reliable communication. Large buffer, wide interconnects and faster router clock (compared to PE clock) are widely used techniques to achieve the goal. However, the proposed hardware SNN is highly resistant to latency. In a typical SNN, the spiking activities is sparse and sporadic \cite{cardin2009driving}. The sparsity is even more visible in the hardware design due to the time-multiplexed nature of the neuron cores. As we will explain in section \ref{sec:operation}, for a neuron core of $M$ logical neurons with $N$ axons, each neuron will be evaluated once every  $(M+C+4)×(N+1)$ cycles. This interval is referred as neuron evaluation cycle (NEC). Assume all neuron evaluations are randomly ordered, the average latency between the spike generation and required spike reception is $T_{NEC}/2$. Furthermore, because of the asynchronous behavior of neurons, it is not absolutely necessary for a spike generated in current NEC to be received in the very next NEC. The STDP window is usually set to be multiple NECs. A communication delay of 1 NEC will hardly affect learning and inference at all. Finally, in-hardware learning automatically helps to adjust synaptic weight based on the hardware. Links that consistently have long latency or dropped packet will eventually have low synaptic weight and hence become less important. Therefore, in this work, our application specific NoC design will aim at minimum silicon area and low overhead.

\begin{figure}[h]
  \centering
  \includegraphics[width=\linewidth]{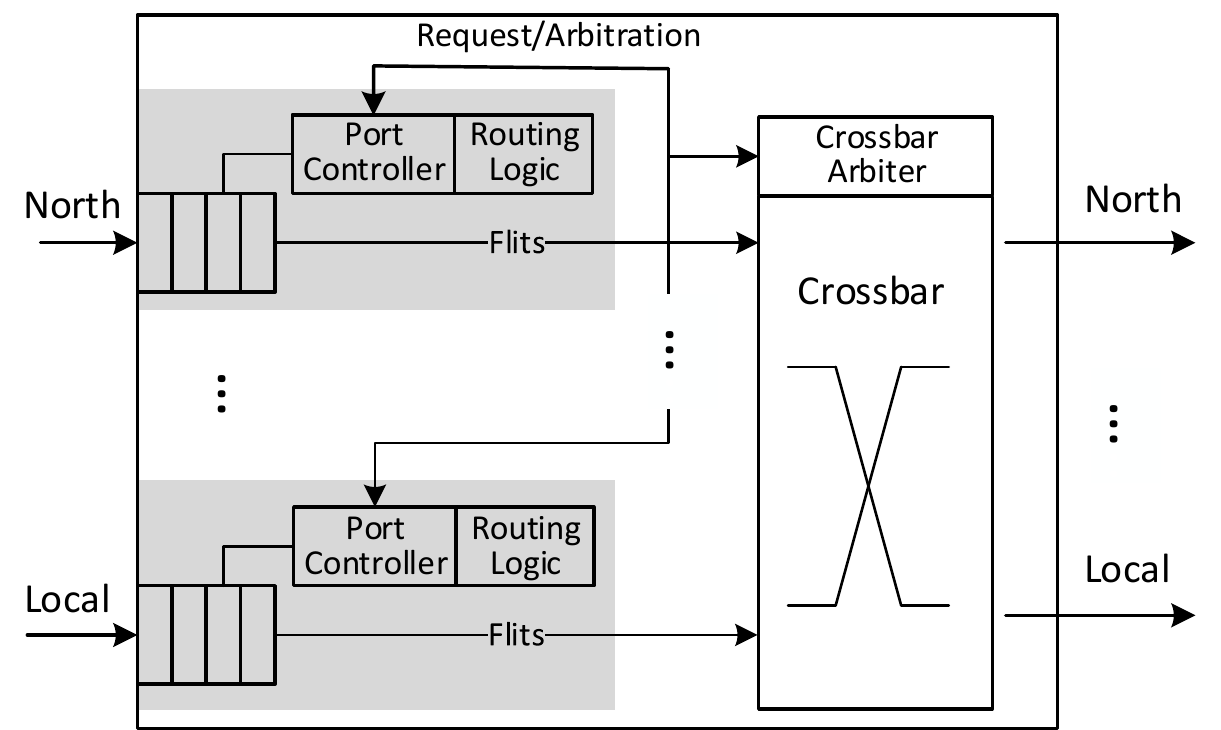}
  \caption{Router Structure}
  \label{fig:router}
\end{figure}

The router consists of five ports, dual clock FIFO, a crossbar switch and an arbiter as shown in Figure \ref{fig:router}. Every port is independent from other ports and all of them work in parallel. Each port has a controller and a routing logic. The routing logic implements routing algorithm and determines the next hop of a packet. The controller detects channel status and coordinates with arbiter to make transmission decisions. The arbiter handles crossbar conflict during the time when an output is requested by multiple inputs. Each router is connected to 4 neighbor nodes, which are north, south, east, west respectively. Each router also connect to its local PE. Each port is attached with a FIFO buffer that can hold one packet. We set the FIFO size to minimum to reduce hardware cost. Routers work at the same frequency as hardware neuron. All routers form a 2-D mesh.

Physical link width is a key factor to the NoC performance and hardware cost. Link is realized by a number of parallel wires connecting two neighboring routers. A wider physical link can provide a larger bandwidth and reduce transmission latency. However, the area overhead of router increases quadratically as the link width increases \cite{lee2013we}. Since the proposed hardware SNN is resistant to latency, we adopt the minimum cost solution and set the link width to be 4 bits.


\subsection{Routing and flow control}

We adopt X-Y routing for its low hardware cost and as its deadlock free \cite{glass1994turn}. Each node holds its own coordinate $(Xc, Yc)$ and the packets contain destination node's coordinate $(Xd, Yd)$. Router compares its own coordinate with the coordinate of the destination and decides where to forward the packet. Horizontal direction has higher priority than vertical direction. If $Xd > Xc$, packet is forwarded to east neighbor, otherwise to the west neighbor. If $Xd = Xc$, packet is routed vertically based on comparison result between $Yc$ and $Yd$. If $Xd = Xc$ and $Yd = Yc$, then packet is forwarded to the local port.

Each packet is an address event representation (AER) consisting of two fields; header and body. The header is $H$ bits absolute coordinates of destination, where $H$ is determined by the NoC size. The body again is divided into two parts. The first part is L-bit axon index, where $2^L=N$, and $N$ is the number of axons of a neuron core. The second part is reserved for debug and function extension.

Router employs wormhole switching as the flow control mechanism. A packet is split into a few flow control units (flits). Each flit has the same size as the physical link width, which is 4-bit. The $H/4$ header flits contain the routing information. As long as header is received, the router can forward the header flits to the next desired hop and all subsequent payload flits will follow the same path. In this way, the asynchronous buffer does not have to store entire packets and both the depth and width of buffer can be minimized mitigating high silicon area cost of the asynchronous buffers. Router stalls transmission when its neighbor is busy.

\subsection{Network Interface}

Network interface(NI) is the bridge between router and Neuron. NI is also responsible for decoding packets and buffering incoming spikes. NI has a register array of length $N$. Each bit corresponds to an input of a hardware neuron. Once a packet is received, the axon ID field is decoded into an address. Specified bit of the register array is set to 1, indicating a spike is received. NI also provides local traffic bypass. All packets targeting the local neuron will be directly decoded.

\subsection{Neuron operations}\label{sec:operation}

Inspired by IBM TrueNorth\cite{merolla2014million}, we designed the hardware neuron architecture aiming at low overhead. In-hardware STDP learning capability is added. The proposed hardware supports two major functions, inference and learning. The inference function integrates weights, updates membrane potential and generates spikes. And the learning function updates weights and bias based on the rules proposed in section \ref{stdp_rules}.

In order to improve resource utilization and achieve high density neurons in SNN, the hardware works in a time-multiplexed manner. The data path and control logic can be used for multiple neurons' computation. We refer to the physical circuit that implements neuron behavior as a physical neuron. Each physical neuron can implement $M$ logical neurons. As we will show in Section \ref{sec:conclusion}, the value of $M$ will affect the speed, cost and energy efficiency of the system.

\begin{figure}[h]
  \centering
  \includegraphics[width=0.5\linewidth]{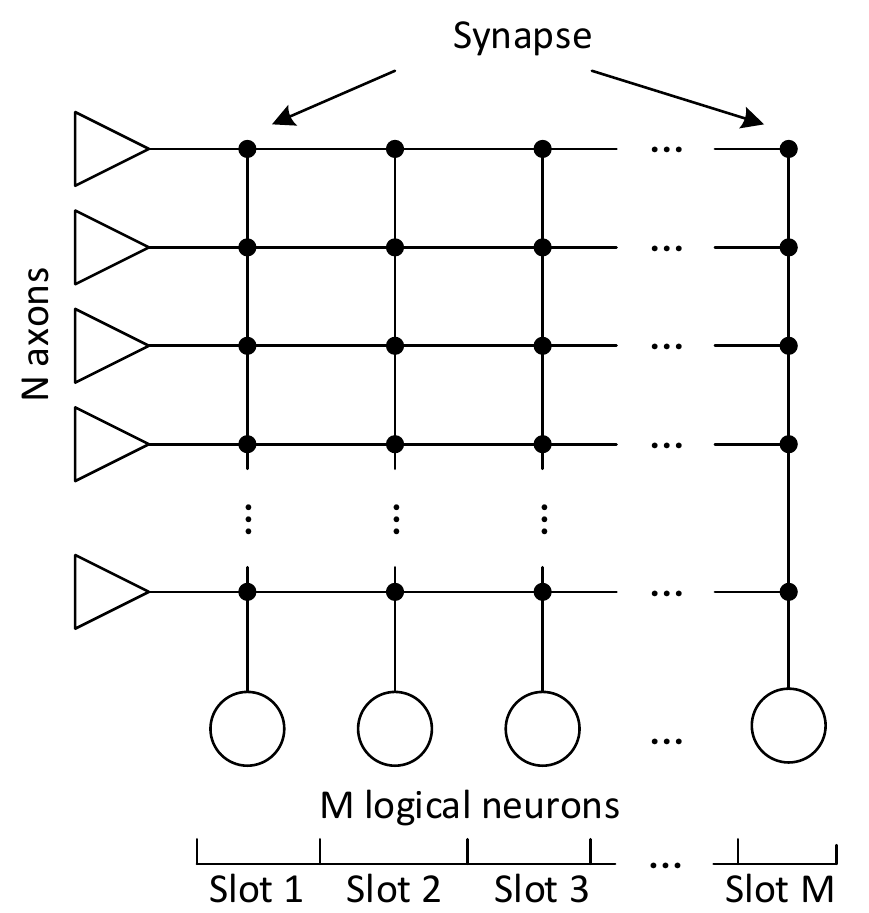}
  \caption{Logical neuron connectivity.}
  \label{fig:crossbar}
\end{figure}

A physical neuron has $N$ inputs called axons. The set of $N$ axons are shared by all logical neurons in a neuron core. In this way, a logical neuron can connect up to $N$ logical neurons through a single spike packet, which reduces the NoC traffic. We refer every connection between an axon and a logical neuron as a synapse. The connectivity between axons and logical neurons can be represented as a crossbar as shown in Figure \ref{fig:crossbar}. Each dot in the figure is a synapse. Every synapse has a unique weight. If a neuron is not connected to an axon, the corresponding synaptic weight is 0.

Each logical neuron performs inference followed by learning, if its learning function is enabled. Learning can only happen after inference, because it requires information such as the firing condition, pre-synaptic history and post-synaptic history, which are updated during the inference. For performance efficiency, we parallelize the learning of the $i^{th}$ neuron with the inference of the $(i+1)^{th}$ neuron.

A global synchronization signal coordinates all physical neurons computation and the interval between two synchronization signals is one NEC. Each neuron is evaluated once every NEC. One NEC is partitioned into $M + 1$ slots; one for each logical neuron and the last slot is to complete the learning operation of the last logical neuron. Each slot has multiple clock cycles and its duration is determined by the learning and inference latencies. As we can see, by evenly distributing logical neurons in multiple physical neurons, less slots are required in a NEC, the whole network can work at a higher frequency.

\subsection{Hardware neuron design}\label{sec:hardware_design}
\begin{figure}
  \centering
  \includegraphics[width=\linewidth]{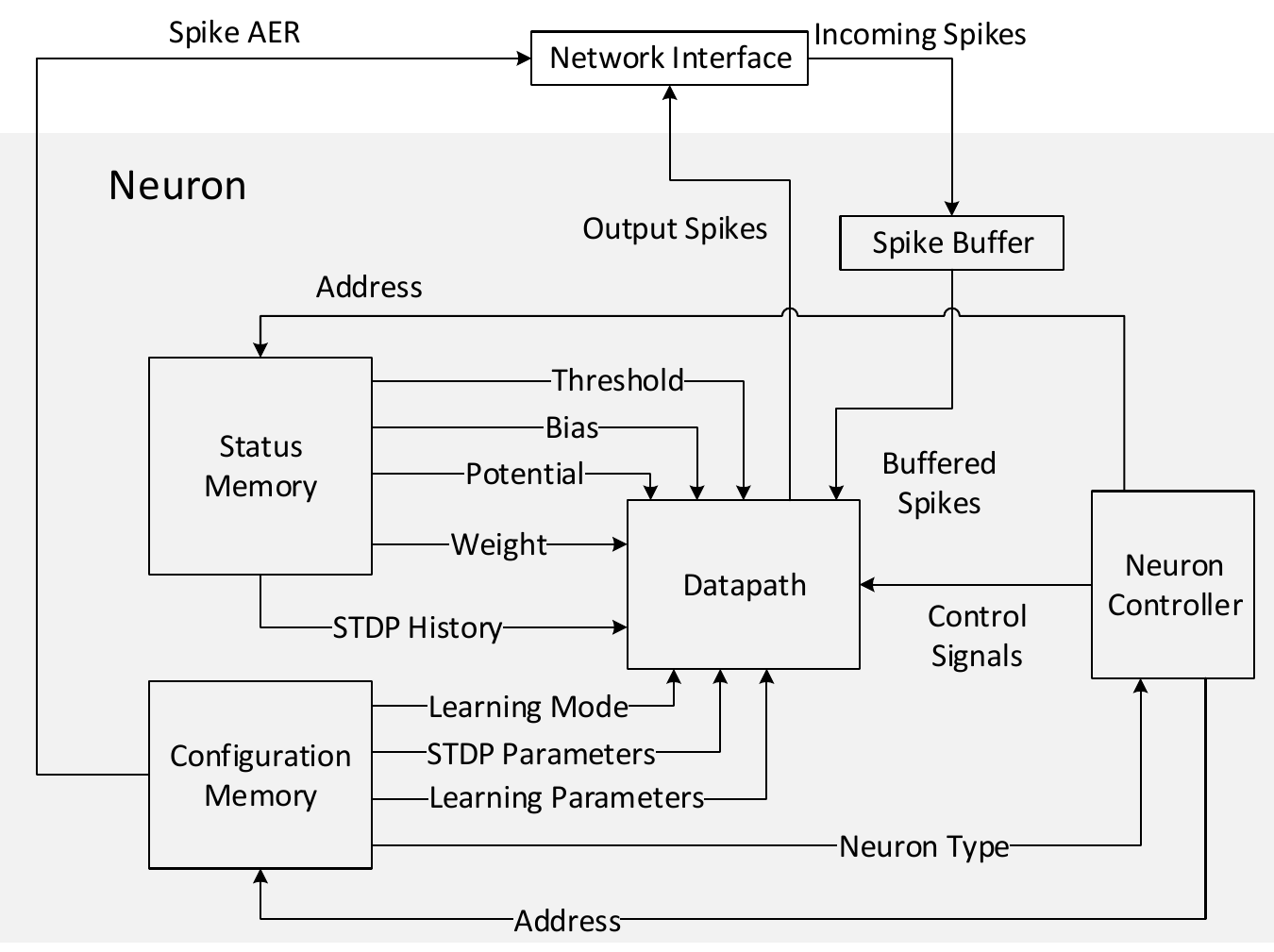}
  \caption{Physical neuron structure}
  \label{fig:neuron_structure}
\end{figure}

As shown in Figure \ref{fig:neuron_structure}, a physical neuron has 5 parts. The neuron controller is responsible for scheduling the computation of logical neurons, generating addresses and control signals for learning and inference. Data path is the key component to implement neuron behaviors, including inference and learning functions. Spike buffer has a register array of length $N$. Each bit corresponds to an axon input. When the start signal is high, the content in spike buffer will be either cleared or overwritten by the output of NI.

There are two memory banks in a hardware neuron.
Configuration memory stores every logical neurons' behavior parameters and learning parameters such as logical neuron type, learning mode, LTP learning rate $(\eta_{LTP})$, LTD learning rate $(\eta_{LTD})$ etc. Status memory stores logical neurons' status parameter, including membrane potential, bias, threshold, axon weights, pre-synaptic history and post-synaptic history. Every physical neuron has its own memory, which is located next to the data path.

The overlapped learning and inference both require to access weight memory at same time. To solve this issue, a FIFO is used. When the ith neuron is performing inference function and accessing weight memory, each weight is pushed into the FIFO. When the inference of $i^{th}$ logical neuron is done, $(i + 1)^{th}$ logical neuron starts inference. At the same time, the learning function of $i^{th}$ logical neuron starts, all required weights of $i^{th}$ are fetched from FIFO and sent to the data path.

A physical neuron with $M$ logical neurons and $N$ axons has $M*N$ synapses, and each synapse has a unique weight. Therefore, weight consumes the most memory resources. The Q2PS STDP rule limits weights in a small range but requires high precision\cite{shrestha2017stable}. This enables reducing integer bits without accuracy penalty. However, some specialized networks such as Winner-take-all circuit\cite{shrestha2017stable} require wide weight range while the precision is not important. To mitigate this problem, each  axon is associated with a scaling factor. By configuring scaling factor, axon can be switched between different precision and range levels, so that wide weight range and high precision can both be satisfied.

Based on its inference function, a logical neuron can be configured as one of the four modes: integrate and fire(IF), stochastic integrate and fire(SIF), spiking Rectified Linear Unit(ReLU) and learning mode. In IF mode, neuron implements Equation 1, if the membrane potential exceeds its threshold, it will forward an AER to router. In SIF mode, the threshold is a random number uniformly distributed in a given range, so that neuron fires at certain probability. In ReLU mode, neuron implements equation \ref{eq:relu}. If learning is enabled, neuron performs the learning rules described in section \ref{stdp_rules}. Separate hardware is used to implement the data path for inference and learning functions.

The inference data path is responsible for computing membrane potential, issuing spikes and performing stochastic behavior. It takes one clock cycle to accumulate the weight of each synapse. Adding the bias and previous NECs membrane potential takes another 2 clock cycles. Comparing current membrane potential to the threshold and determining whether to spike or not takes one more cycle. At last the new membrane potential is written back to status memory. Assuming the axon number is configured as $N$, the inference evaluation of 1 logical neuron takes $N+4$ clocks.

\begin{figure}[t]
  \centering
  \includegraphics[width=\linewidth]{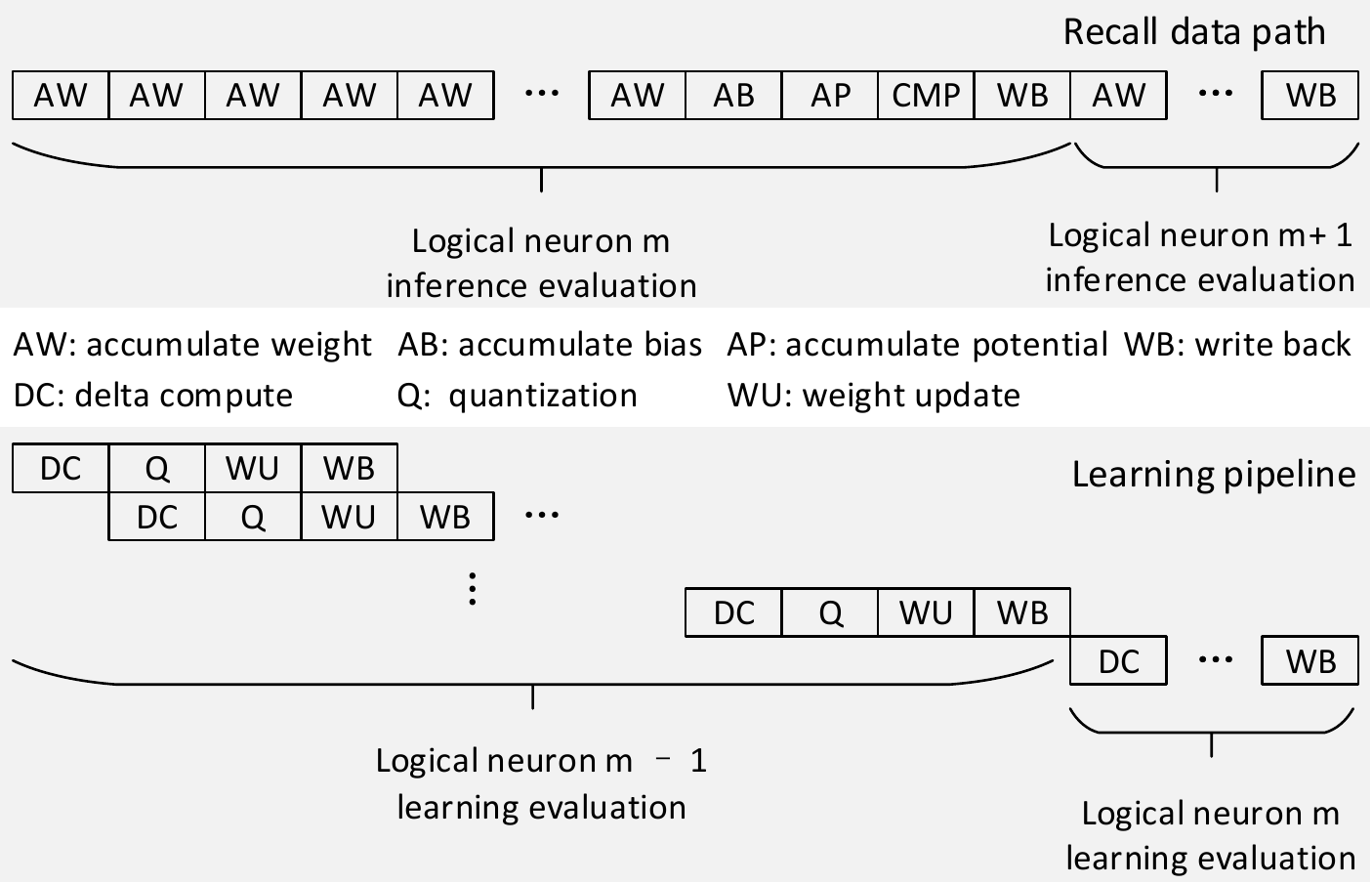}
  \caption{Data path pipeline}
  \label{fig:pipeline}
\end{figure}

The learning data path implements the Q2PS rule, which uses adder, shifter, priority encoder and look-up table(LUT) to approximate the exponential STDP learning. Learning data path is pipelined and has four stages. In stage one, based on the spiking history and spiking condition, $\eta_{LTD}$ or $\eta_{LTP}$ is selected to perform LTP or LTD learning. Q is computed as equation \ref{eq:ltp} or equation \ref{eq:ltd}. In stage two, $\bar{Q}$ is obtained by priority encoding $Q$. Then equation 3 is computed, 1 is shifted by $\bar{Q}$ times to get the change of synaptic weight. In stage three, weight change is applied to the old weight. In stage four, updated weight is written to weight memory. The learning of bias is also implemented in this stage. The difference between bias learning and weight learning is that the bias learning is not a function of time, whether LTD or LTP is used depends only on the current spiking condition. Learning of a neuron with $N$ axons also takes $N + 4$ clocks. Figure \ref{fig:pipeline} shows the timing of data path and pipeline.

In addition to the data path for inference and learning, an STDP tracker is implemented to maintain the pre-synaptic spike history and post-synaptic spike history which are critical to performing correct learning activity. The post-synaptic history tracker is a counter that is set to 0 in the NEC when the logical neuron generates a spike, and incremented by 1 in every NEC otherwise. The pre-synaptic history tracker is also a counter that is set to 0 in the NEC when a spike is received on that synapse, and incremented by 1 otherwise. Post-synaptic/pre-synaptic history valid flag is asserted if the the tracker is less than the LTD/LTP window.

When learning mode is enabled, STDP tracker determines to expire post-synaptic history and pre-synaptic history based on the firing condition, valid flag and incoming spike. When the logical neuron issues a spike and pre-synaptic history is valid, STDP tracker expires pre-synaptic history. When logical neuron receives a spike, and post-synaptic history is valid, STDP tracker will expire post-synaptic history.

The proposed hardware also supports multicast. In the multicast mode, the most significant bit of extension field in spike AER is a control bit. If the MSB is 1, neuron controller will increase address pointer to read next spike AER from configuration memory and keeps sending write request to NI until a spike AER's MSB is 0. In this way, a logical neuron can have flexible number of destinations, which allows the network to support more complex topologies, simplify configuration and improve resource utilization.

\section{RESULTS AND DISCUSSION}
The proposed hardware design is implemented in Verilog RTL-level model and synthesized on Altera Arria 10 platform. The results in terms of learning and performance of the NoC design is discussed, and the design space is explored in this section.

\subsection{Unsupervised Learning of MNIST digits}\label{sec:mnist}

The stochastic firing and STDP learning enables unsupervised feature learning in SNNs. To validate the functioning of the hardware design, we employ a simple pattern learning task. In this task, we utilize a simple winner-take-all (WTA) circuit to learn handwritten digits 0 and 1\cite{ahmed2016probabilistic} from MNIST data set. The network is trained using 100 samples and each sample is exposed to the network for 100 NECs. In this experiment we set $M$ and $N$ to be 128 and 256 respectively.

For convenience, given the size of fan-in for a core (256 axons), we look to reduce the required number of inputs into any layer. As an MNIST image has 28x28 pixels, we employ an average-pooling-like mechanism for patches of 2x2 with a stride of 2 in the first layer. Thus the resultant input into the second layer will be an average-pooled 14x14 MNIST image. The overall network consists of two layers. The first layer consists of 196 average pooling neurons whose fan-in is 2x2, and the second layer consists of 4 SIF neurons with STDP learning enabled and 4 ReLU neurons to relay the spikes. These 8 neurons form a winner-takes-all (WTA) circuit as described in \cite{ahmed2016probabilistic}. The 196 neurons in the first layer are mapped to 14 cores and all neurons in the second layer are mapped 1 core. A 4x4 mesh network is configured to perform this experiment. MNIST image is encoded into spike packets and a dedicated router is used to inject external packets.

\begin{figure}
  \centering
  
    \begin{subfigure}{\linewidth}
    \includegraphics[width=\linewidth]{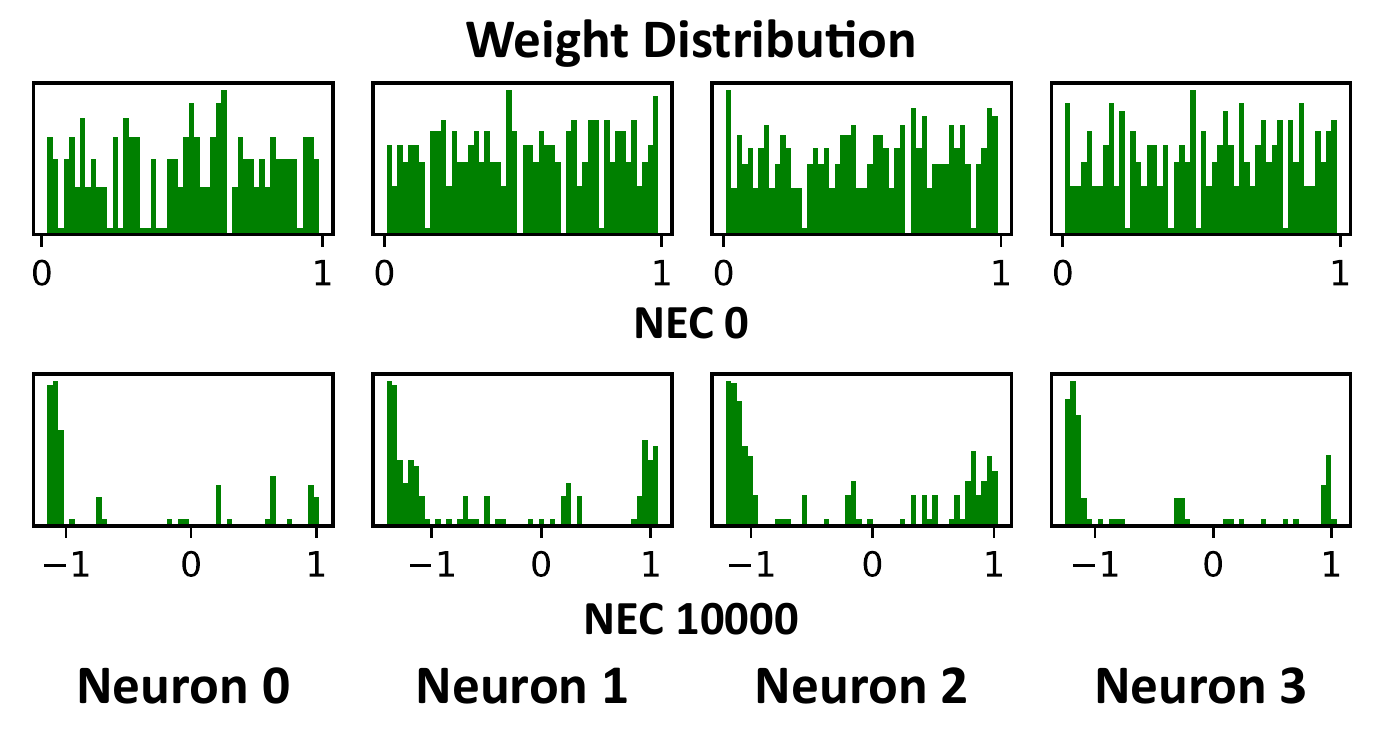}
    \caption{Weight distribution.}\label{fig:weight_distribution}
  \end{subfigure}
  
  \begin{subfigure}{\linewidth}
    \includegraphics[width=\linewidth]{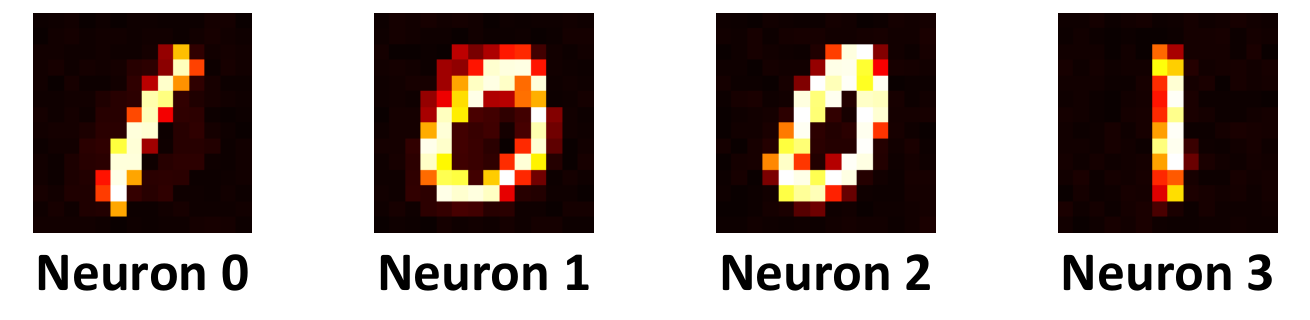}
    \caption{Weight visualization}\label{fig:weight_visualization}
  \end{subfigure}
  
  \caption{Learnt weights}
  \label{fig:spike_activity}
\end{figure}

Figure \ref{fig:weight_distribution} shows the distribution of the weight before and after learning. The stability of the Q2PS STDP rule can be seen from this figure. As all learnt weights are limited in range [-1.41 , 1.07], less integer bits are used to encode the weights and hence lower memory usage. The selectivity provided by the Q2PS STDP rule can also be observed. Before the learning, the weight follows a uniform distribution in the range [-1, 1], while after learning the diverging weights of the network form a bimodal distribution as expected in \cite{shrestha2017stable}. Figure \ref{fig:weight_visualization} gives the weight map of the 4 classification neurons. As we can see that each of them learned specific patterns.

\begin{figure}
  \centering  
    \begin{subfigure}{\linewidth}
    \includegraphics[width=\linewidth]{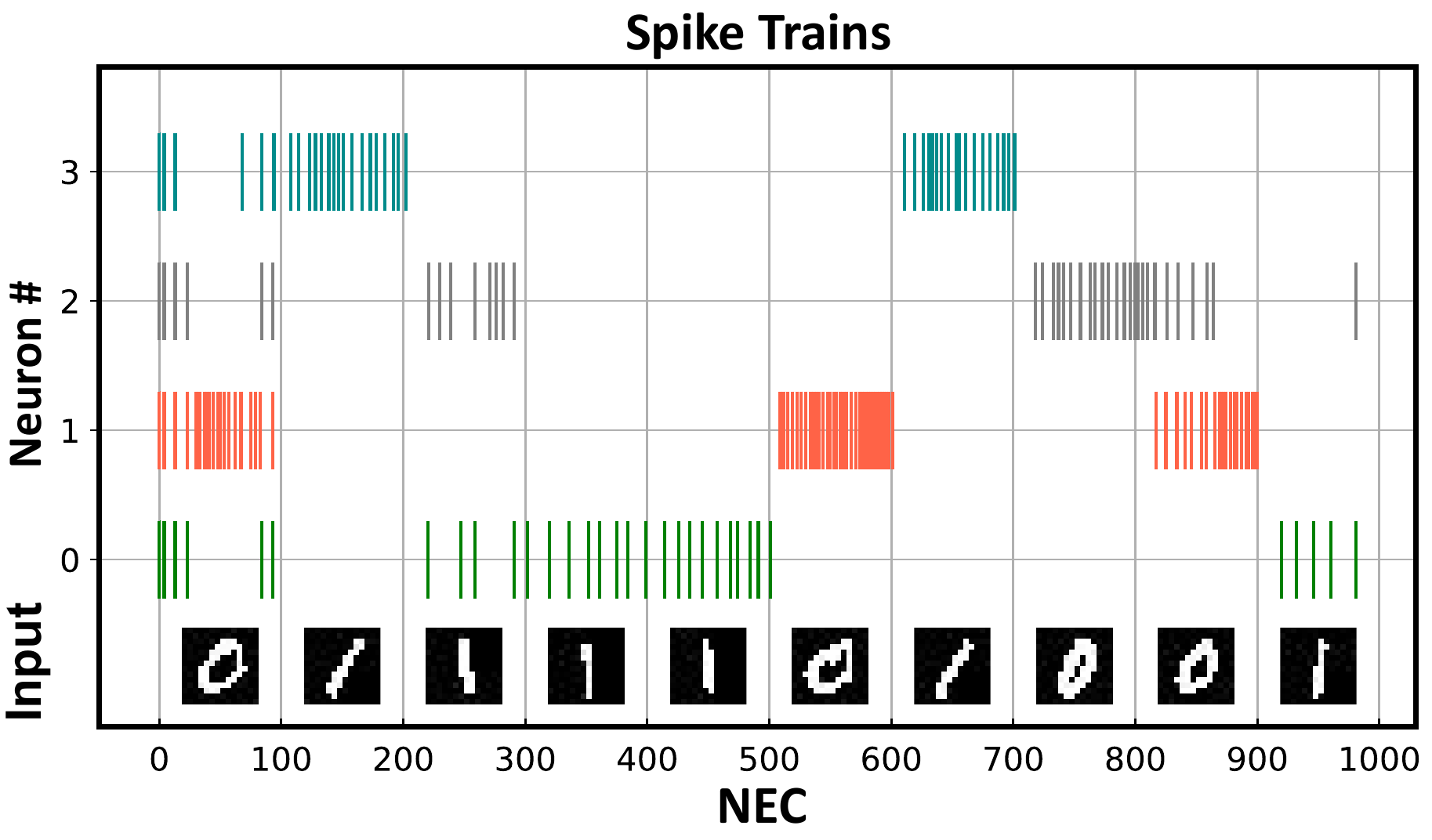}
    \caption{Spike train pattern and corresponding input.}\label{fig:spike_train}
  \end{subfigure}
 
  \begin{subfigure}{\linewidth}
    \includegraphics[width=\linewidth]{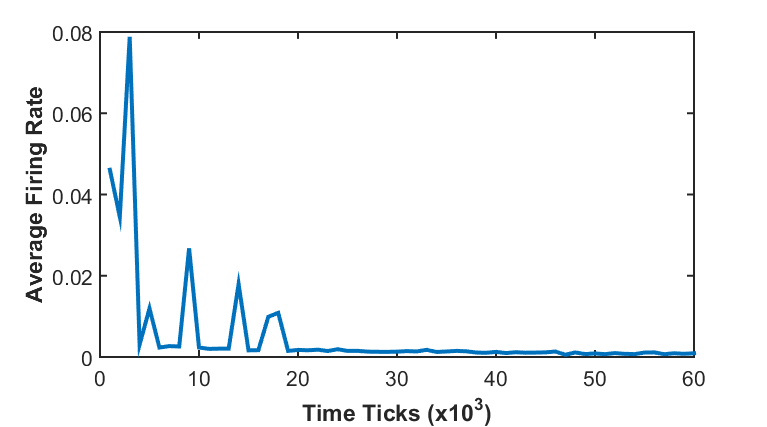}
    \caption{Average firing rate of learning neurons.}\label{fig:avgfiringrate}
  \end{subfigure}
  
  \caption{Learning neuron spike activity}
  \label{fig:spike_activity}
\end{figure}

Figure \ref{fig:spike_train} gives the learning results. Initially, the spiking activity is random, but as the time goes on, a spiking pattern emerges which represent the corresponding selectivity of the neurons. This selectivity is also visible from the firing rates of the learning neurons shown in Figure \ref{fig:avgfiringrate}. Initially the spiking rate is high as all the neurons are randomly firing. As time goes on, the neurons start learning patterns and only fire selectively. Thus, drastically reducing the firing rate.

Most neurons remain idle through the entire training and the average firing probability is 11.472\%. Because the computation is event-driven, the sparsity of neuron activation leads to low power consumption.

\subsection{Performance evaluation}\label{sec:performance}

The traffic pattern and spike pattern of NoC are specific to the topology and neuron parameters of SNNs that are mapped to it. In order to guarantee the proposed architecture can satisfy the requirements of various applications, we perform a pressure test to evaluate the performance of NoC by increasing firing rate using above the MNIST recognition network a baseline. We configured a 4x4 mesh network with random connectivity. Each physical neuron has $M = 128$ logical neurons and $N = 256$ input axon. 2048 logical neurons are distributed in 16 physical neurons. The router has buffer depth of 2 packets. Table \ref{tab:traffic} shows NoC traffic statistics under different firing rates. The first row is the above MNIST network that is used as baseline. Each experiment runs 1000 NECs. At the operation frequency of 200 MHz, $5 * 10^7/(257*260) = 1,489$ NECs can be executed in 1 second. Benefiting from the sparse activation of SNNs and shared input axon, the traffic stays at a relatively low level. When firing rate is 87.526\%, the NoC achieves throughput of 26 Mbyte/s. The forth column in Table \ref{tab:traffic} shows the average latency of the network under different firing rate. The MNIST baseline has smaller latency because it has shorter average packet route than random topology. The MNIST baseline also has larger traffic because external input of image is injected to network. The average latency remains stable under different traffic load. All packet can be delivered in current NEC, which is 33560 clock cycles in this case. We observe no packet drop due to congestion. The experiment shows that, although small flit size cause large delay, there is still sufficient time to deliver packets.

\begin{table}
\centering
\caption{Traffic Statistic}
\label{tab:traffic}
\begin{tabular}{|c|c|c|c|c|c|c|}
\hline
\textbf{\thead{Firing \\ Rate}} & \textbf{\thead{Total \\ Spikes}} & \textbf{\thead{Traffic \\ (byte) }} & \textbf{\thead{Avg. \\Latency \\ (Cycle)} }& \textbf{\thead{Max \\ Latency}} & \textbf{\thead{Min \\latency}} \\
\hline
11.513\%    & 218196       & 9324660      & 33.328     &151 & 23    \\
\hline
10.723\%    & 220495       & 2219028       & 50.780    &99 &  23      \\
\hline
15.911\%    & 327165       & 3297500       & 50.906    &96&  23     \\
\hline
39.940\%    & 821245       & 8266052       & 51.369   &110&   23   \\
\hline
46.995\%    & 966316       & 9849392       & 51.806    &101&  23    \\
\hline
54.633\%    & 1123366      & 11080148      & 50.983    &99&   23   \\
\hline
62.743\%    & 1290114      & 12644556      & 50.964    &108&  23    \\
\hline
72.609\%    & 1492974      & 14988016      & 51.907    &106&   23   \\
\hline
81.393\%    & 1673593      & 16657080      & 51.876   &108& 23       \\
\hline
87.562\%    & 1800452      & 17906592      & 51.842   &106&   23\\
\hline
\end{tabular}
\end{table}

\begin{figure}
  \centering
  \includegraphics[width=\linewidth]{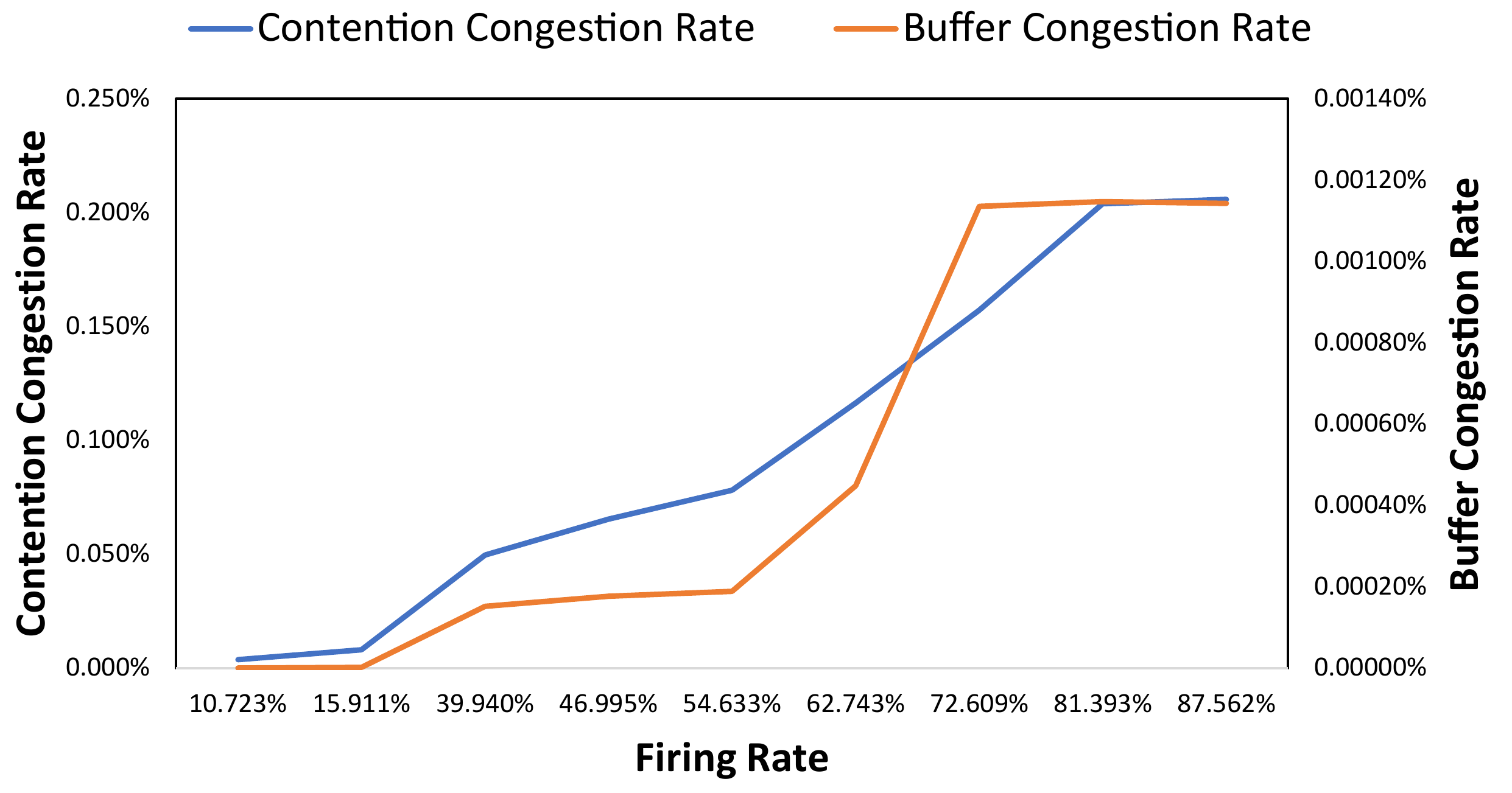}
  \caption{Congestion rate}
  \label{fig:congestion_rate}
\end{figure}

Two types of congestion are studied. The first type of congestion occurs inside the router, which is caused by multiple incoming packets requesting the same destination port. In this case, only one port is granted to transmit while other ports are stalled temporarily until the transmission is done. We refer to this as contention congestion. The second type is buffer congestion, which occurs when the destination routers input buffer is full, router has to wait until the packets stored in the destination are transmitted. A packet will be dropped when buffer congestion occurs. We define the congestion rate as $T_{congestion} / T_{execution}$, where $T_{congestion}$ is the total clock cycles in which congestion has occurred, and $T_{execution}$ is the running time of the entire simulation. As shown in Figure \ref{fig:congestion_rate}, both the two types of congestion rate increase as the firing rate increases. However, due to the time-multiplexed design, a physical neuron can generate 1 spike every 260 clocks at most, the traffic is spanned in a long duration and at most times routers remain idle. The sparsity of SNN activation makes the traffic even sparser. As a result, even in the worst case, where the network has a firing rate of 87.562\%, the congestion is still rare.

In SNNs, neurons firing rates are about 10\%\cite{cardin2009driving}. There is no significant performance degradation observed in the worst case, therefore the proposed architecture is able to satisfy the requirements of various applications.

\subsection{Design space exploration}\label{sec:design_suggestion}
The physical neuron capacity, NoC size and router buffer size can affect power consumption, parallelism and efficiency. Here, we study the impacts of these factors and provide guidelines for the design of hardware spiking neural network.

First, there is a trade-off between physical neuron capacity and parallelism. Assume that a SNN has 2048 neurons and a physical neuron can contain 256 logical neurons. 2048 / 256 = 8 physical neurons are required to map the SNN to hardware. In this case, a NEC should have at least (256 + 1) * 260 = 66820 clock cycles. If the physical neuron can contain 32 neurons, 2048 / 32 = 64 physical neurons are required. A NEC should contain (32 + 1) * 260 = 8,580 clock cycles. Therefore the second hardware SNN runs approximately 8 times faster than the first one as the second hardware SNN require less clock cycles in a NEC. However this improvement is obtained at the cost of silicon area and power consumption.

Table \ref{tab:neuron_capacity} shows the impact of physical neuron capacity on FPGA resource consumption, power and energy when mapping a SNN with 2048 neurons to hardware. The experiment is performed at the operating frequency of 200 MHz. LUT consumption has an approximately linear relation to the number of physical neuron number. The logic resource consumption of a cell is almost constant. Memory consumption increases slightly as the physical neuron number grows. Extra memory consumption is introduced by router's input buffer. The last column shows the energy consumption of executing 1000 NECs. Although power is significantly larger when more physical neurons are required, the length of a NEC becomes shorter. Hence more NECs can be executed in a given running time. For FPGA implementation, it is preferable to use small physical neuron to increase parallelism as well as energy efficiency. 

\begin{table}
\centering
\caption{Impact of Physical Neuron Capacity}
\label{tab:neuron_capacity}
\begin{tabulary}{0.8\linewidth}{|c|c|c|c|c|c|c|}
\hline
\textbf{Size}   & \textbf{NEC}   & \textbf{Neuron} & \textbf{LUT}   & \textbf{Memory} & \textbf{Power}   & \textbf{Energy}    \\ 
                & (cycle)         & \textbf{number}         &                 & (byte)         & (mW)                 & (J)       \\
\hline
32       & 8580      & 64     & 18.97\% & 1,394,688  & 5759.05   & 0.247     \\
\hline
64       & 16900     & 32     & 9.56\% & 1,359,872  & 2144.83   & 0.351     \\
\hline
128      & 33540     & 16     & 4.87\% & 1,342,464  & 1839.52   & 0.498     \\
\hline
256      & 66820     & 8      & 2.46\% & 1,333,760  & 1679.96   & 0.783     \\
\hline
\end{tabulary}
\end{table}

Another factor that has impacts on performance is router's buffer size. Larger buffer size can reduce congestion rate, however the dual clock buffer is expensive in area. It is desirable to reduce router area to improve neuron density. We studied the impact of buffer size on network performance. A 4x4 network is configured, which consists of 2048 neurons. Two sets of experiments are performed. In the first set, the SNN's firing rate is approximately 10\%, which is close to the firing rate of realistic applications. In the second set, a pressure test is performed, the SNN has approximately 100\% firing rate. Network performance are shown in table \ref{tab:buffer}.
Compared with buffer of depth 8, buffer congestion decreases considerably, contention congestion also decreases due to less buffer congestion. The buffer congestion is rare when buffer depth is 16. Increasing buffer further does not bring significant performance improvement.

\begin{table}
\centering
\caption{Impact of router buffer size}
\label{tab:buffer}
\begin{tabulary}{\linewidth}{|l|l|l|l|l|}
\hline
\textbf{Firing rate} & \textbf{Buffer size}   & \textbf{Contention}  & \textbf{Buffer}     \\ 
       & (width*depth)    & \textbf{congestion}       & \textbf{congestion}       \\
\hline
10.251\%    & 4*8              & 0.0331\%         & 0.037\%      \\
\hline
10.251\%    & 4*16             & 0.0328\%         & 0\%          \\
\hline
10.251\%    & 4*32             & 0.0328\%         & 0\%              \\
\hline
99.896\%    & 4*8              & 2.891\%          & 4.002\%         \\
\hline
99.896\%    & 4*16             & 2.625\%          & 0.015\%         \\
\hline
99.896\%    & 4*32             & 2.625\%          & 0\%             \\
\hline
\end{tabulary}
\end{table}

\section{Conclusions} \label{sec:conclusion}
In this paper, we presented a comprehensive system-level spiking neural network hardware implementation, which features scalability, flexibility and in-hardware STDP learning capability. The proposed design is validated by unsupervised MNIST digits learning.

\bibliographystyle{IEEEtran}
\bibliography{DAC2018_hardware}
%




\end{document}